\documentclass[twocolumn,prl,tightenlines,superscriptaddress,showpacs]{revtex4-1}

\usepackage{amsmath}
\usepackage{amssymb,amsfonts,latexsym}
\usepackage[mathcal]{euscript}
\usepackage{graphicx}
\usepackage{epsfig}
\usepackage{xfrac}

\begin{document}

\title{Collective Motion of Self-Propelled Particles with Memory}

\author{Ken H. Nagai}
\affiliation{School of Materials Science, Japan
Advanced Institute of Science and Technology, Ishikawa 923-1292, Japan}

\author{Yutaka Sumino}
\affiliation{Department of Applied Physics, Tokyo University of Science, Tokyo 125-8585, Japan}

\author{Raul Montagne}
\affiliation{Departamento de Fisica, UFRPE, 52171-900, Recife, PE, Brazil}

\author{Igor S. Aranson}
\affiliation{Materials Science Division, Argonne National Laboratory,  Argonne, Illinois 60439, USA}

\author{Hugues Chat\'{e}}
\affiliation{Service de Physique de l'Etat Condens\'e, CNRS UMR 3680, CEA-Saclay, 91191 Gif-sur-Yvette, France}
\affiliation{LPTMC, CNRS UMR 7600, Universit\'e Pierre et Marie Curie, 75252 Paris, France}
\affiliation{Beijing Computational Science Research Center, 3 Heqing Road, Beijing 100080, China}

\date{\today}

\begin{abstract}

We show that memory, in the form of underdamped angular dynamics, is a crucial ingredient for 
the collective properties of self-propelled particles. Using Vicsek-style models with an Ornstein-Uhlenbeck
process acting on angular velocity, we uncover a rich variety of collective phases not observed in usual overdamped
systems, including vortex lattices and active foams. In  
 a model with strictly nematic interactions the smectic arrangement of Vicsek waves giving rise to global polar order
is observed. 
We also provide a calculation of the effective interaction between vortices in the case where
a telegraphic noise process is at play, explaining thus the emergence and structure of the vortex lattices observed here and in 
motility assay experiments. 
\end{abstract}

\maketitle

Self-propelled particles are nowadays commonly used to study collective motion and more generally ``dry" active matter, 
where the surrounding fluid is neglected. Real world relevant situations
include shaken granular particles \cite{KUDROLLI,NARAYAN,aranson2007,DYCOACT,KUMAR}, 
active colloids \cite{PALACCI,BOCQUET,BRICARD}, 
bio-filaments displaced by motor proteins \cite{SCHALLER,SUMINO,DOGIC}. 
The trajectories of moving living organisms (from bacteria to large animals such as fish,  birds and even human crowds) are also routinely modeled
by such particles, see e.g. \cite{BACT-SPP,WENSINK,MIDGES,BIRD,FISH,HUMAN}. 

Many of these `active particles' travel at near-constant speed with their dynamics modeled as a persistent random walk with some 
stochastic component acting directly on their orientation \cite{ABP}. This noise, which represents external and/or internal perturbations, produces 
jagged irregular trajectories. Most of the recent results on active matter 
have been obtained in this context of overdamped dynamics. 

In many situations, however, the overdamped approximation is not justified. In particular,
trajectories can be essentially smooth, as for chemically propelled rods \cite{paxton2004,takagi2013}, birds, some large fish that swim steadily \cite{FISH}, or even biofilaments in 
motility assays with a high density of molecular motors \cite{SUMINO}. Whether underdamped dynamics can make a difference
at the level of collective asymptotic properties is largely unknown.
Interesting related progress was recently reported for starling flocks \cite{CAVAGNA}. Underdamped ``spin" variables are instrumental there
for efficient, fast transfer of information through the flock, allowing swift turns in response to threats during which speed is
modulated in a well coordinated manner.
In the other examples cited above, speed remains nearly-constant and the persistently turning tracks of fish or microtubules reveal some
finite, possibly large, memory of the curvature. In this context an Ornstein-Uhlenbeck (OU) process acting on the {\it angular velocity}
was shown to be a quantitatively-valid representation \cite{SUMINO,FISH}.
The collective motion of self-propelled particles with 
such underdamped angular dynamics 
remains largely unknown.

In this Letter, we explore minimal models of aligning self-propelled particles with  memory 
similar to that used in \cite{SUMINO} to study the emergence of large-scale vortices in motility assays.
Considering both polar (ferromagnetic) and nematic alignment interactions between particles, we
show that memory is a crucial ingredient for collective motion giving rise to a wealth of
collective states heretofore not observed in the memory-less case. 
Among the most remarkable features
of the rich phase diagram of these minimal models, we provide evidence that global {\it polar} order can arise
from strictly {\it nematic} interactions, taking the form of trains of ``Vicsek waves", i.e. the ubiquitous 
nonlinear structures well-known from models and experimental situations described by the traditional Vicsek model \cite{VICSEK,GREGOIRE}.
We not only consider OU, but also telegraphic noise (TN) processes which we show
to be more easily amenable to analytic approaches. 
For these telegraphic noise models, we present a calculation yielding the effective interaction
between emerging vortices, and thus a simple explanation of why these structures form hexagonal or square lattices
depending on the symmetry of the alignment interaction.

Let us first introduce Vicsek-style models with memory. Point particles with position $\dot{\bf x}_i$ move at unit speed along 
their heading $\theta_i(t)$, i.e. $\dot{\bf x}_i= {\bf e}_{\theta_i}$, where 
${\bf e}_{\theta_i}$ is the unit vector along $\theta_i$. 
Headings evolve according to:
\begin{equation}
\label{eq1}
\frac{{\rm d}\theta_i}{{\rm d}t} = \frac{\alpha}{{\cal N}_i} \sum_{|{\bf x}_j-{\bf x}_i |<1} \sin [m(\theta_j - \theta_i)]  \;+\; \omega_i(t)
\end{equation} 
where the sum is over the ${\cal N}_i$ neighbors of $i$ within unit distance,
$m=1$ (resp. 2) codes for ferromagnetic (resp. nematic) alignment,
and $\omega$ is a zero-mean noise. 
For uncorrelated white noise, noise strength and global density
of particles are the two main parameters, and these overdamped models are expected to exhibit a phase diagram similar to that of their discrete-time counterparts \cite{POLAR,NEMA,RODS}.

Memory is introduced via an Ornstein-Uhlenbeck process for the noise $\omega$ in  Eq. \eqref{eq1}, keeping all other compartments of the dynamics overdamped:
\begin{equation}
\label{eq2}
\frac{{\rm d}\omega_i}{{\rm d}t} = -\frac{1}{\tau} \omega_i + \xi_i(t)
\end{equation}
where $\xi$ is a Gaussian white noise of variance $\sigma^2$.
The underdamped models defined by Eqs.~(\ref{eq1}),(\ref{eq2}) depend on one extra parameter,
the memory time $\tau$. In the limit of small $\tau$, they reduce to their
overdamped versions.

In \cite{SUMINO}, the nematic version ($m=2$) of this model was introduced and a preliminary study of its collective 
regimes was presented in the $(\rho_0,\tau)$ plane keeping 
$\langle\omega^2\rangle=\frac{1}{2}\tau\sigma^2$, the variance of $\omega$, fixed~\cite{NOTE1}.
Here, we present a detailed phase diagram of our models 
in the  $(\rho_0,\tau)$ plane also keeping $\langle\omega^2\rangle$ fixed, having checked that 
its global structure does not vary much with $\langle\omega^2\rangle$. 
The quantifiers used to define and distinguish the collective states observed
and the methodology followed to establish the following phase diagrams are detailed in \cite{EPAPS}.

\begin{figure}[t!]
\includegraphics[width=\columnwidth]{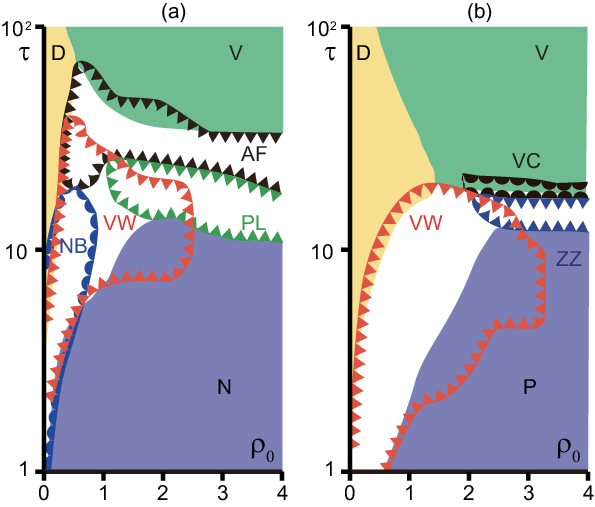}
\caption{(Color online) Phase diagrams of the models defined by Eqs.~\eqref{eq1},\eqref{eq2}. 
See main text and/or Fig.~\ref{fig2} for definitions of the various phases.
(a): nematic model.
(b): polar model.
}
\label{fig1}
\end{figure}

\begin{figure*}[t!]
\includegraphics[width=\textwidth]{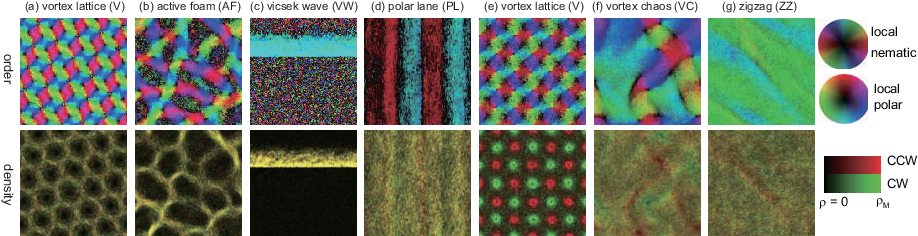}
\caption{(Color online) Various phases for the nematic (a-d) and polar (e-g) model. Corresponding movies are available in \cite{EPAPS}.
Top: local orientational order (nematic (a,b), polar (c-g)), color stands for orientation, intensity for modulus.
Bottom: superimposed density of CW ($\omega<0$, green) and CCW ($\omega>0$, green) particles.
Colormaps are on the right, with  $\rho_M=2$ for (a-e), $4$ for (f) and $5$ for (g).
Parameters ($\rho_{0}$, $\tau$): (a)(1,100), (b)(1,40), (c)(0.56,10), (d)(1.78,20), (e)(1,100), (f)(4.22, 17.5), and (g)(3.16, 15). 
Periodic square of linear size 512, $\alpha = 0.1$, $\langle\omega^2\rangle^{1/2} = 0.03$  (resp. $0.06$) for the nematic (resp. polar) model.
}
\label{fig2}
\end{figure*}

{\it Nematic model} ($m=2$). Its phase diagram  is summarized in Fig.~\ref{fig1}a. 
The three main regions found in \cite{SUMINO} are  present: 
At small enough density, noise always dominates alignment leaving an homogeneous disordered phase ({\sf D}). 
At low enough $\tau$, homogeneous nematic order emerges ({\sf N}), 
while at high enough $\tau$ one eventually observes a hexagonal lattice of large vortices ({\sf V}) made with local nematic order
(see Fig.~\ref{fig2}a, and \cite{SUMINO}).

Our analysis also revealed the existence of a complex arrangement of other collective states in the area separating these
three main regions. 
Consistent with  the overdamped case \cite{RODS}, 
a coexistence phase made of a dense nematic band ({\sf NB})
standing in a disordered gas is present between the {\sf D} and {\sf N} regions. 
But there is also a large domain inside which one can observe trains of dense, traveling bands such as those
known to occur in overdamped models with ferromagnetic alignment like the original Vicsek model \cite{POLAR,BANDS}. 
These smectic polar pattern (`Vicsek waves', {\sf VW}, hereafter, Fig.~\ref{fig2}c) are even the {\it only} asymptotic state in
a subregion of their observation domain.
Thus global {\it polar} order arises whereas the interaction is purely nematic. \cite{NOTE4}
Note that such global order, made of nonlinear structures, is {\it not} in contradiction with
the theoretical arguments at linear level \cite{Aranson2013,MCM-RODS,RODS} precluding the emergence of homogeneous polar order in systems with nematic interactions.

The vortex lattice directly melts to the homogeneous disorder state when decreasing $\tau$ at low densities. 
For large-enough $\rho_0$, on the other hand, it 
becomes a spatio-temporally disordered cellular structure ---an `active foam' ({\sf AF}) (Fig.~\ref{fig2}b).
This foam is still made of the same nematically-ordered streams as the vortex lattice, but they are unstable.

\begin{figure}[t!]
\includegraphics{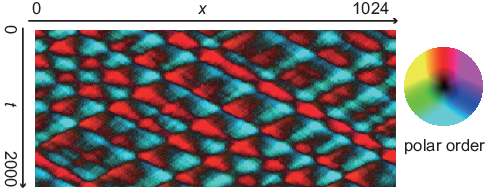}
\caption{(Color online) Space-time diagram  in the laning phase (nematic model).
Rectangular domain $(L_x, L_y)= (1024 \times 256)$ with global nematic order along $y$ (periodic boundary conditions).
Local polar order averaged over $y$ is represented (colormap on right). 
Parameters: ($\rho_{0}$, $\tau$)=(1.33,20) (other parameters as in Fig.~\ref{fig2} for nematic model)
}
\label{fig3}
\end{figure}

Finally, at large enough densities, increasing $\tau$, the homogeneous nematic state becomes transversally unstable.
Global nematic order is preserved, but local segregation leads to oppositely-going polar lanes of relatively large 
width (laning region {\sf PL} in Fig.~\ref{fig1}a, snapshot in Fig.~\ref{fig2}d). 
Laning has been reported in overdamped self-propelled hard rods with long-enough aspect ratio interacting exclusively 
via steric exclusion, and also for elongated `deformable' particles \cite{LANING1,LANING2}. Observed here it with pointwise particles, 
we conclude that finite-size particles are not necessary.  
This striped pattern is dynamic: a space-time plot of local polar order
across the lanes reveals {\it two} sets of lanes (each with lanes in both orientations) 
moving slowly along this transversal direction, yielding a standing wave pattern
(Fig.~\ref{fig3}). Whether laning is also dynamic in rods systems remains to be investigated.

{\it Polar model} ($m=1$). 
We now describe the phase diagram in the case of ferromagnetic alignment (Fig.~\ref{fig1}b), 
which has the same general features as in the nematic case: Three main regions are present, 
homogeneous disorder ({\sf D}), homogeneous polar order ({\sf P}), and vortex lattice ({\sf V}). This lattice is now 
a checkerboard arrangement of locally polar clockwise and counterclockwise vortices (Fig.~\ref{fig2}e). 
Again, for short-enough memory time $\tau$, one recovers, between the {\sf D} and {\sf P} phases, 
the coexistence phase made of a smectic arrangement of Vicsek waves ({\sf VW}), familiar from the study of the overdamped case.
The central region is less complicated than in the nematic case, but nevertheless comprises (at least) two new phases. 
For $\rho_0>1$, the vortex lattice leaves place to chaotic phase where 
vortex cores move, coalesce, vanish, and form spontaneously (``vortex chaos" {\sf VC}, Fig.~\ref{fig2}f). 
Finally, the homogeneous polar order region {\sf P} is bordered, at large $\tau$, by a regime 
characterized by moving domains inside which polar order is perpendicular to the main direction of motion, 
so that particles perform ``zigzag" trajectories ({\sf ZZ} region in Fig.~\ref{fig1}b, snapshot in Fig.~\ref{fig2}g).

Some general comments are in order. For both the nematic and polar models several states are observed to coexist
in the central region of parameter space: They 
can be reached from different initial conditions, and observed for at least the very long time $T$  that we use here as a criterion for ``stability". 
Determining which one is dominant, though, is a difficult task beyond the scope of this paper. 
However, as already noticed for Vicsek waves in the nematic case, there exists, for each phase, 
a region where it is the {\it only} observed state. Thus all phases reported here are likely to exist asymptotically. 

{\it Telegraphic noise models}.
Further progress may come from numerical work even more intensive than reported here.
Another avenue to ascertain the robustness of our results would be to derive continuous theories for our models. 
Unfortunately, methods that were 
proven successful in the overdamped case
are difficult to apply here since they rely on decorrelation of particles between successive interactions. 
A way out of this conundrum is to replace 
the Ornstein-Uhlenbeck process \eqref{eq2} by some symmetric Poisson telegraphic noise process  $\omega=\epsilon \omega_0$
in which $\epsilon = \pm 1$  with a switching probability equal to $1/\tau$. 
Then, indeed, one can consider two subpopulations of particles, clockwise and counterclockwise, exchanging 
particles at rate $1/\tau$ via some {\it memoryless} Poissonian switching process, similar to that for flipping self-propelled rods in \cite{takagi2013}. 

We first studied numerically the collective properties of the telegraphic noise (TN) models. 
Their phase diagrams are qualitatively similar to those presented in Fig.~\ref{fig1} for the OU models
(detailed results will be published elsewhere \cite{TBP}).
Almost all phases observed in the OU case 
are also present with TN. Possibly the only exception is the absence of the active foam regime, 
which is replaced by a chaotic regime that appears as a superposition of nematic bands and vortices. 
This points to the robustness of the results presented so far, an indication that Vicsek-style models with memory 
can probably all be described by some common continuous theory. 

Here, we postpone the construction 
of a full-fledged hydrodynamic theory. Instead, we now show how the structure of the emerging vortex lattice
observed at large $\tau$ and the shape of its region of existence in parameter space 
can be understood via the calculation of the effective interaction between vortices.

\begin{figure}[t!]
\includegraphics[width=\linewidth]{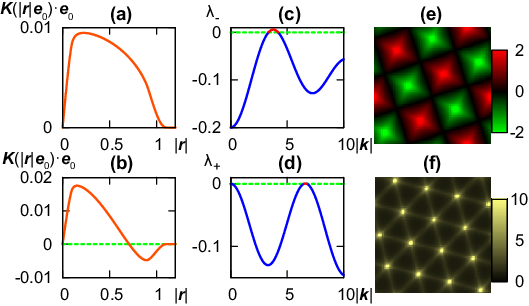}
\caption{(Color online) Continuous description for vortex lattice formation. Top (bottom) row: polar (nematic) model.
(a,b) effective interaction;
(c, d) dispersion relation (eigenvalue for $F_{+}-F_{-}$, $F_{+}+F_{-}$);
(e,f) steady states when $\rho_{0} \tau$ is larger than the critical value. (e) shows
 $F_{+}-F_{-}$, and (f) shows $F_{+}+F_{-}$.  Periodic square of linear
 size $40$, $D=10$, $\alpha=0.05$, and $\rho_{0}=2$. In the
  polar (nematic) interaction case, $\rho_{0}\tau=20$ ($17$).
}
\label{fig4}
\end{figure}

In the vortex region, $\tau$ is large, which means that one
particle typically stays on a circle trajectory for a while. 
We take advantage of this to describe the dynamics of the {\it centers} ${\bf r}_{i}$ of these circular trajectories, 
changing ${\bf x}_{i}$, the particle position,  to ${\bf r}_{i} ={\bf x}_{i} + \epsilon_i \omega_0^{-1} (-\sin\theta_i,\cos\theta_i)$. 
Differentiating both sides and using Eq. \eqref{eq1} we obtain
\begin{equation}
 \dot{{\bf r}}_{i}  \!=\!
\frac{2\epsilon_{i}\delta(t\!-\!t_{\rm s})}{\omega_0} \tilde {\bf e}_{\theta_{i}} \!-\!
 \frac{\alpha \epsilon_i }{{\cal N}_i\omega_0}  \! \! \sum_{|{\bf x}_i-{\bf x}_j|<1}
 \sin [m (\theta_{j} \!-\! \theta_{i})]   {\bf e}_{\theta_{i}} 
\label{120947_28Oct14}
\end{equation}
where  $\tilde  {\bf e}_{\theta_i} = {\bf e}_{\theta_{i} -  \frac{\pi}{2}}$, $t_{\rm s}$ is the time when $\epsilon_{i}$ switches between $\pm 1$.
Assuming that ${\cal N}_i=1$, $\theta$ changes little during one
collision, we describe  the ensemble behavior of Eqs. (\ref{120947_28Oct14}) by probability distributions  of clock/anticlock-wise particles 
$f_\pm({\bf r}, \theta) $  satisfying the 
equations (see \cite{takagi2013})
\begin{eqnarray}
\frac{\partial f_\pm({\bf r}, \theta)}{\partial t }  &=&  \pm \frac{\partial}{\partial \bf r } \!\cdot\!   {\bf e}_\theta  f_\pm({\bf r}, \theta) \frac{U_{\pm}}{\omega_{0}} \! -\!  
\frac{\partial}{\partial \theta  }     f_\pm({\bf r}, \theta) \left(\pm \omega_{0} \!+\! U_{\pm}\right)\nonumber  \\
 &&  + \frac{f_\mp({\bf r}\pm D \tilde{\bf e}_{\theta}, \theta ) - f_\pm({\bf r} , \theta) }{\tau} 
\label{fpe}
\end{eqnarray}
with the ensemble-averaged rotation rates 
\begin{eqnarray}
U_{\pm} \!=\! \alpha \!\!\!\sum _{\mu =-1,1} \int{\!\rm d}{\bf \xi}\! \int_{-\pi}^{\pi}   \!\!\! {\rm d}\phi 
 \chi ({\bf x}^{\mu}_{{\bf \xi},\phi},
 {\bf x}^{\pm 1}_{{\bf r},\theta} ) f_{\mu }({\bf \xi},\phi) 
 \sin(\phi - \theta)  
 \nonumber 
\end{eqnarray}
Here, $\chi({\bf x}^{\prime},{\bf x})$ is $1$ ($0$) when $|{\bf
 x}^{\prime} - {\bf x}|<(\geq)1$, ${\bf
 x}^{\epsilon}_{{\bf r},\theta}={\bf r}+\epsilon/\omega_{0} \tilde{\bf e}_{\theta_{i}}$, and $D=2/\omega_{0}$.
To simplify Eqs. (\ref{fpe}), we average over the  rotation period and assume the distributions to be almost uniform in $\theta$ 
(an assumption valid below the onset of ordered phase  \cite{IGOR-LEV}). 
Defining $F_{\pm } (\bm{r}) = \int_{0}^{2 \pi} f_\pm({\bf r}, \theta) {\rm d} \theta/2 \pi   $
 we obtain
 \begin{align}
 \frac{\partial F_{\pm}}{\partial t} &= -\!D \alpha
 \frac{\partial}{\partial {\bf r}} \!\cdot\!\! F_{\pm}({\bf r}) \!\! \int \!\! {\rm d}{\bf \xi}
  {\bf K}({\bf r}
 \!-\! {\bf \xi}) [F_{\pm}({\bf \xi}) + \!(-1)^{m}\! F_{\mp}({\bf \xi})]\notag\\
 &  + \frac{1}{\tau} \left\{\frac{1}{2\pi}\int_{0}^{2\pi} \!\! {\rm d}\theta
 F_{\mp}\left({\bf r} + D {\bf e}_{\theta}\right)  - F_{\pm}({\bf r})  \right\},\label{eqF}
 \end{align}
where the radial interaction kernel
\begin{equation}
{\bf K}({\bf r})= -\! \int _{-\pi}^\pi \! {\rm d} \theta \int
 _{-\pi}^\pi \! {\rm d} \phi \chi({\bf x}^{1}_{{\bf 0},\phi}, {\bm
 x}^{1}_{\bf r,\theta})  \sin[m (\phi -\theta)] \frac{{\bf e}_\theta}{4 \pi} 
\end{equation}
is  shown  in Fig.~\ref{fig4}a,b. (see \cite{EPAPS} for  details). 
The interaction is short-range: ${\bf K}({\bf r})=0$
for $|{\bf r}|>D+1$.
In the polar case ($m=1$), it is attractive (repulsive) between vortices
with the same (opposite) sign of $\omega$.
In the nematic case ($m=2$), the interaction has a short-range attractive part and a medium range 
repulsive part.

We examined the stability of a homogeneous  state $F_+ = F_- = \rho_{0}/2$. 
Linearizing Eqs.~\eqref{eqF}, we obtain after Fourier transform 
\begin{align}
 \frac{{\rm d}F_{\pm}^{\bm{k}}}{{\rm d} t} = &\left((-1)^{m+1} \pi D \alpha |\bm{k}|
 K^{|\bm{k}|} \rho_{0} + \frac{ J_{0}(|\bm{k}|D)}{\tau}     \right)
 F^{\bm{k}}_{\mp}\notag\\
 &-\left(\pi D \alpha |\bm{k}|
 K^{|\bm{k}|} \rho_{0} + \frac{1}{\tau}
					     \right)F^{\bm{k}}_{\pm}.\label{eqFlin}
\end{align}
$J_{n}$ is the $n$-th Bessel function of the first kind, and
$K^{k}= \int_{0}^{D+1}{\rm d}r \ J_{1}(kr)r \bm{e}(0)\cdot \bm{K}(r\bm{e}(0))$. 
Examining  the eigenvalues of Eq.~\eqref{eqFlin}, we find that the homogeneous state becomes  unstable when $\rho_{0}\tau>c_{\rm min}$,   
$c_{\rm min} = \min _{|\bm{k}|}
\frac{(-1)^{m}J_{0}(|\bm{k}|D)-1}{2\pi D\alpha |\bm{k}| K^{|\bm{k}|}}$
 (Fig.~\ref{fig4}c,d).
In the nematic case, due to the conservation of particles,  the most unstable mode $F_{+} + F_{-}$ is  neutral at $|\bm{k}|=0$, see Fig.~\ref{fig4}(d).  
The corresponding amplitude equation 
is the conserved Swift-Hohenberg equation
 \cite{elder}. It yields hexagonal lattices, consistent with our findings. 
In the polar case,  the unstable mode $F_+ -  F_-$  would yield the conventional Swift-Hohenberg equation with cubic nonlinearity. 
But conservation of particles implies  that 
the unstable mode couples to the neutral mode $ F_+ + F_-$  at $|\bm{k}|=0$. A square lattice is then expected \cite{oscillons}, again in agreement
with our results. 
Numerical integration of Eqs.~\eqref{eqF}  confirms that 
for $\rho_0 \tau> c_{\rm min}$  the correct periodic lattice emerges,  Fig.~\ref{fig4}e,f.
The instability condition $\rho_{0}\tau>c_{\rm min}$ is in semi-quantitative agreement with
simulations of the  particle models with TN.

To summarize, we have shown that memory, in the form of underdamped angular dynamics, is a crucial ingredient for 
the collective properties of self-propelled particles. Our Vicsek-style models exhibit a prominent 
vortex lattice phase at high density and large memory. With nematic alignment an hexagonal lattice emerges, but polar 
interactions lead to a checkerboard square lattice of alternating  vortices, a fact explained by the 
calculation. 
We have also shown the emergence of a number of collective states for moderate memory times. 
In particular, we observed the emergence, out of purely nematic
interactions, of global, long-range, polar order made of Vicsek waves. 
We also reported that nematic interactions can give rise to a system of traveling polar lanes. 
Such laning configurations have been reported for overdamped self-propelled elongated objects 
interacting via steric exclusion \cite{LANING1} . Thus repeated collisions between these rods seem to 
amount to some effective memory. Laning may in fact be best characterized this way, and not 
necessarily by the aspect ratio of particles.

Even though smooth trajectories with persistent curvature are observed for  birds, 
swimmers, etc., the studied  models  are probably most suited for self-organization phenomena 
in {\it in vitro} mixtures of biofilaments and motor proteins such as motility assays. 
Nematic vortices were already reported in \cite{SUMINO}, and 
Vicsek wave-like patterns in \cite{SCHALLER}. We are confident that some of the other phases reported here are also 
present in such systems. 

\acknowledgements
We thank the Max Planck Institute for the Physics of Complex Systems, Dresden,
for providing the framework of the Advanced Study Group
``Statistical Physics of Collective Motion'' within which part of this work was conducted. 
K. H. N. was supported by a Grant-in-Aid for Scientific Research on Innovative Areas 
``Fluctuation \& Structure'' (No.26103505), and a JSPS fellowship for
young scientists (No. 23-1819).   Y. S. was supported by a
Grant-in-Aid for Young Scientists B (No. 24740287).
I. S. A.  was supported by the US Department of Energy, Office of Science, Basic Energy Sciences,
Materials Science and Engineering Division.


\begin{thebibliography}{99}

\bibitem{KUDROLLI} A. Kudrolli, {\it et al.}, Phys. Rev. Lett. {\bf 100}, 058001 (2008).

\bibitem{NARAYAN} V. Narayan, S. Ramaswamy, N. Menon, Science {\bf 317}, 5834 (2007).
\bibitem{aranson2007} I.S. Aranson, D. Volfson, and L. S. Tsimring, Phys. Rev.  E {\bf 75}, 051301 (2007).

\bibitem{DYCOACT} J. Deseigne, O. Dauchot, and H. Chat\'e,
Phys. Rev. Lett. {\bf 105}, 098001 (2010);
J. Deseigne, S. L\'eonard, O. Dauchot, and H. Chat\'e, Soft Matter, {\bf 8}, 5629 (2012).

\bibitem{KUMAR}  N. Kumar, H. Soni, S. Ramaswamy, and A.K. Sood,
Nature Comm.  {\bf 5},  4688 (2014).

\bibitem{PALACCI} J. Palacci, {\it et al.}, Science {\bf 339}, 6122 (2013).

\bibitem{BOCQUET} I. Theurkauff, C. Cottin-Bizonne, J. Palacci, C. Ybert, and L. Bocquet, Phys. Rev. Lett. {\bf 108}, 268303 (2012).

\bibitem{BRICARD} A. Bricard, {\it et al.}, Nature {\bf 503}, 95 (2013).

\bibitem{SCHALLER}
V. Schaller {\it et al.}, Nature {\bf 467}, 73 (2010);
Soft Matter, {\bf 7}, 3213 (2011);
Proc. Natl. Acad. Sci. USA {\bf 108}, 19183 (2011).

\bibitem{SUMINO} Y. Sumino, {\it et al.}, Nature {\bf 483}, 448 (2012).

\bibitem{DOGIC} T. Sanchez {\it et al.}, Nature {\bf 491}, 441 (2012).

\bibitem{BACT-SPP} F. Peruani, {\it et al.}, Phys. Rev. Lett. {\bf 108}, 098102 (2012).

\bibitem{WENSINK} H. H. Wensink, {\it et al.}, Proc. Natl. Acad. Sci. (New York) {\bf 109}, 14308 (2012).

\bibitem{MIDGES}  A. Attanasi, {\it et al.}, Phys. Rev. Lett. {\bf 113}, 238102 (2014).

\bibitem{BIRD}  H. Hildenbrandt, C. Carere, and C.K. Hemelrijk, Behavioral Ecology {\bf 21}, 1349 (2010);
D.J.G. Pearce {\it et al.}, Proc. Natl. Acad. Sci. (New York) {\bf 111}, 10422 (2014);
A. Cavagna, {\it et al.} preprint arXiv:1403.1202 (2014);

\bibitem{FISH} J. Gautrais, {\it et al.}, J. Math,. Biol. {\bf 58}, 429 (2009);
PLoS Comp. Biol. {\bf 8}, e1002678 (2012); J. Theor. Biol. {\bf 58}, 429 (2009).

\bibitem{HUMAN} I.  Karamouzas, B. Skinner, and S. J. Guy,  Phys. Rev. Lett. {\bf  113},   238701 (2014)

\bibitem{ABP} P. Romanczuk, {\it et al.}, Eur. Phys. J. Special Topics {\bf 202}, 1 (2012).

\bibitem{paxton2004} W.F. Paxton et al. JACS 2004;{\bf 126} 13424 (2004)
\bibitem{takagi2013} D. Takagi, A. B. Braunschweig, J. Zhang, and M. J. Shelley, Phys. Rev. Lett. {\bf 110}, 038301 (2013) 

\bibitem{CAVAGNA} A. Attanasi, {\it et al.}, Nature Phys. {\bf 10}, 692 (2014);
A. Cavagna, {\it et al.} preprint arXiv:1410.2868 (2014).

\bibitem{GREGOIRE} G. Gr\'egoire and H. Chat\'e, Phys. Rev. Lett. {\bf 92}, 025702 (2004).

\bibitem{VICSEK}  T. Vicsek {\it et al.},
Phys. Rev. Lett. {\bf 75}, 1226 (1995).

\bibitem{POLAR}
H. Chat\'e, {\it et al.}, Phys. Rev. E {\bf 77}, 046113 (2008); 
A. Solon, H. Chat\'e, and J. Tailleur, preprint arXiv:1406.6088 to appear in Phys. Rev. Lett. (2015).

\bibitem{NEMA} H. Chat\'e, F. Ginelli, and R. Montagne, Phys. Rev. Lett. {\bf 96}, 180602 (2006);
S. Ngo, {\it et al.}, Phys. Rev. Lett. {\bf 113}, 038302 (2014). 

\bibitem{RODS} F. Ginelli, {\it et al.}, Phys. Rev. Lett. {\bf 104}, 184502 (2010); 
A. Peshkov, {\it et al.}, Phys. Rev. Lett. {\bf 109}, 268701 (2012).

\bibitem{NOTE1}  This choice guarantees that for large $\tau$ the size of  vortices,  
which is governed by $\langle\omega^2\rangle$, remains roughly constant.

\bibitem{EPAPS} See EPAPS Document No. XXX.

\bibitem{BANDS} The Vicsek model bands first reported in \cite{GREGOIRE} have since been observed in many other models. See, e.g. 
S. Mishra, A. Baskaran, and M.C. Marchetti, Phys. Rev. E {\bf 81}, 061916 (2010);
C.A. Weber, {\it et al.}, Phys. Rev. Lett. {\bf 110}, 208001 (2013);
P. Romanczuk and L. Schimansky-Geier, Interface Focus {\bf 2}, 746 (2012).

\bibitem{NOTE4}  When starting from random initial conditions, counter-propagating bands are typically observed. After a 
long transient, one direction eventually wins, leaving a wave train  moving in the same direction --- global polar order.


\bibitem{Aranson2013}  C.W. Harvey, M.  Alber, L.S.  Tsimring, and I.S.  Aranson,  New Jour. Phys. {\bf  15}, 035029 (2013).

\bibitem{MCM-RODS} A. Baskaran and M.C. Marchetti,
Phys. Rev. E {\bf 77}, 011920 (2008);
Phys. Rev. Lett. {\bf 101}, 268101 (2008).

\bibitem{LANING1}
H. H. Wensink and H. L\"owen, J. Phys. Cond. Mat. {\bf 24}, 464130 (2012);
S.R. McCandlish, A. Baskaran, and M.F. Hagan, Soft Matter {\bf 8}, 2527 (2012);
A. M. Menzel, J. Phys., Cond. Mat. {\bf 25}, 505103 (2013);
T. Gao, {\it et al.}, preprint arXiv:1401.8059 (2014);
H.-S. Kuan, {\it et al.},  preprint arXiv:1407.4842 (2014).

\bibitem{LANING2}
A. M. Menzel and T. Ohta, Europhys. Lett. {\bf 99}, 58001 (2012).

\bibitem{TBP} Y. Sumino, {\it et al.}, to be published.

\bibitem{IGOR-LEV} I.S. Aranson and L.S. Tsimring, Phys. Rev. E {\bf 71}, 050901 (2005);
{\it Ibid.} {\bf 74}, 031915 (2006).
\bibitem{elder} K. R. Elder,M.  Katakowski, M.  Haataja, and M.  Grant, \prl {\bf 88}, 245701 (2002) 
\bibitem{oscillons} L.S. Tsimring and I.S. Aranson, \prl {\bf 79 }, 213   (1997)  

\end{thebibliography}
\end{document}